\documentstyle[prb,twocolumn,aps,graphicx]{revtex}

\begin{document}

\draft

%\preprint{{\number\month}/{\number\day}/{\number\year}}

\twocolumn
[\hsize\textwidth\columnwidth\hsize\csname@twocolumnfalse\endcsname
%\begin{title}
\title{Comment on "Magnetoelastic model for the relaxation of
lanthanide ions in YBa$_2$Cu$_3$O$_{7-\delta}$ observed by neutron
scattering"}
%\end{title}

\author{A.T. Boothroyd$^1$}

%\begin{instit}
\address{
$^1$ Department of Physics, Oxford University, Oxford, OX1 3PU,
United Kingdom. }
%\end{instit}

\date{\today}
\maketitle

\begin{abstract}
Lovesey and Staub have argued \protect{[S.W. Lovesey and U. Staub,
Phys. Rev. B {\bf 61}, 9130 (2000)]} that experimental data on the
temperature dependence of the paramagnetic relaxation of
lanthanide ions doped into YBa$_2$Cu$_3$O$_{6+x}$ are in agreement
with the predictions of a model that describes the relaxation as
due to the scattering of phonons via a magnetoelastic interaction.
By generalising their model I show that the level of agreement is
strongly dependent on the number of intermediate lanthanide energy
levels included in the calculation, and that inclusion of a more
complete set of levels leads to very different results that do not
necessarily support the phonon damping picture.
\end{abstract}

 ]
%\newpage
%\twocolumn
%\narrowtext

In a recent paper \cite{LS-PRB-2000}, Lovesey and Staub considered
how lattice vibrations could contribute to the relaxation of a
transition between the crystal field split energy levels of a
paramagnetic ion. Their work was prompted by a number of recent
publications
\cite{Osborn-PhysicaC-1991,Mukherjee-PRB-1994,Amoretti-PhysicaC-1994,Boothroyd-PRL-1996,Mesot-EPL-1998,Staub-JPCM-1999,Roepke-PRB-1999,Rubio-Temprano-PRL-2000}
reporting neutron spectroscopic measurements of the temperature
dependence of crystal field transition linewidths for various
lanthanide ions doped into cuprate superconductors. Lovesey and
Staub's contention is that the relaxation observed for these
materials is dominated by decay of the crystal field excitation
into lattice vibrations, rather than into spin excitations on the
CuO$_2$ planes as has been assumed by other authors. Lovesey and
Staub arrive at this viewpoint by comparing experimental data with
the results of a model that assumes a magnetoelastic interaction
between the paramagnetic ion and the lattice.

The main purpose of this Comment is to raise awareness of an
approximation made by Lovesey and Staub in calculating the
relaxation from the magnetoelastic coupling model. The
approximation involves the neglect of all but 3 of the crystal
field split $4f$ levels in the relaxation calculation. For the
case of Ho$^{3+}$ impurities in superconducting
YBa$_2$Cu$_3$O$_{6+x}$ I will show that the use of this 3-level
model has led the authors to obtain unreasonably good agreement
between the calculated relaxation and the experimental data, and
that inclusion of other thermally populated lanthanide energy
levels gives results that do not differ significantly from those
calculated with the exchange coupling model used previously. I
will also show that the value of the coupling constant in the
magnetoelastic model derived from the data depends strongly on the
number levels included in the calculation. I will argue,
therefore, that the accord between theory and experiment presented
in Ref. \onlinecite{LS-PRB-2000} cannot be taken as evidence that
phonons rather than spin excitations are the most important source
of relaxation.

I begin by outlining the factors which enter into a calculation of
the relaxation of crystal field transitions. Following Lovesey and
Staub let us consider a transition between two eigenstates
$|a\rangle$ and $|b\rangle$ of the lowest energy $J$ multiplet of
a lanthanide ion in a crystal field. For simplicity let
$|a\rangle$ be the ground state. The lifetime of the $|a\rangle
\rightarrow |b\rangle$ transition is finite because of coupling to
electronic or vibrational degrees of freedom in the host system.
Relaxation from $|b\rangle$ back to $|a\rangle$ can take place
either directly or indirectly, the latter proceeding via an
intermediate lanthanide energy level $|\gamma\rangle$. If
$|\gamma\rangle$ is higher in energy than $|b\rangle$ then for the
indirect transition to take place the $4f$ ion must be able to
absorb energy from the host, and this can be done if the
temperature $T$ is large enough to generate a significant thermal
population of excitations in the host with energies equal to the
difference in energy between $|a\rangle$ or $|b\rangle$ and
$|\gamma\rangle$. The density of host excitation states at the
relevant energy will clearly be relevant in determining the
importance of a given relaxation channel, whether direct or
indirect, and a further factor is the size of the transition
matrix elements connecting the states involved. The relevant
matrix elements for the models under consideration are those of
the quadrupolar operators $Q_{\nu}$ in the case of magnetoelastic
coupling, and of the total angular momentum operators $J_{\alpha}$
in the case of exchange coupling. Here $\nu$ is a symmetry label,
and $\alpha=x,y,z$.

In principle there are as many indirect relaxation channels as
there are intermediate states $|\gamma\rangle$, but because there
often exists a wide variation in the sizes of the transition
matrix elements the effectiveness of each channel varies
considerably. In Ref. \onlinecite{LS-PRB-2000} it was argued that
inclusion of only 3 states ($|a\rangle$, $|b\rangle$ and a single
$|\gamma\rangle$) allows the relaxation calculation to be
simplified while retaining all the essential features.

On the basis of this 3-level approximation Lovesey and Staub
derived an expression (Eq.\ (6.7) of Ref.
\onlinecite{LS-PRB-2000}) for the temperature dependence of the
transition linewidth, and fitted their model to the experimental
linewidth data for Tb$^{3+}$, Ho$^{3+}$, and Tm$^{3+}$ in
superconducting YBa$_2$Cu$_3$O$_{6+x}$ assuming a Debye density of
states to describe the lattice vibrations. The calculation for
Ho$^{3+}$ is reproduced in Fig.\ 1 together with the data from
Boothroyd {\em et al.}\cite{Boothroyd-PRL-1996}. The energy of the
$|a\rangle \rightarrow |b\rangle$ transition is 0.5\,meV, and the
value of the intermediate level $|\gamma\rangle$ used to generate
the calculated curve is 11.8\,meV.

From the quality of the agreement shown in Fig.\ 1 it is tempting
to be satisfied with the use of the 3-level model, and indeed for
the Tb$^{3+}$, and Tm$^{3+}$ systems there are sound physical
arguments based on the energies of the excited crystal field
levels and size of the quadrupole matrix elements to suggest that
the totality of intermediate levels can reasonably be approximated
by a single level. These arguments are discussed in Ref.
\onlinecite{LS-PRB-2000}. In the case of Ho$^{3+}$, however, the
validity of the 3-level approximation is considerably harder to
justify because there are no fewer than 6 crystal field levels in
the energy range 1--12\,meV several of which have substantial
quadrupole matrix elements connecting them to levels $|a\rangle$
or $|b\rangle$. Hence, these levels will contribute to the
measured relaxation of the 0.5\,meV $|a\rangle \rightarrow
|b\rangle$ transition through indirect processes.

These considerations make it worthwhile extending the
magnetoelastic model to include an arbitrary number of
intermediate levels. For simplicity we will take level $|a\rangle$
to be the ground state and level $|b\rangle$ the first excited
state, and re-label these states as $|0\rangle$ and $|1\rangle$
respectively. In this simplification all other intermediate levels
$|\gamma\rangle$ are higher in energy than $|1\rangle$, as occurs
in the experimental systems considered here, but it is
straightforward to treat other sequences of levels.  Let
$\omega_{\gamma}$ be the energy of $|\gamma\rangle$ relative to
the ground state ($\hbar=1$). Generalising the results of Ref.
\onlinecite{LS-PRB-2000} we can express the linewidth (half width
at half maximum) in the magnetoelastic model as
\begin{displaymath}
\Gamma_{\rm ME} = \sum_{\nu}\Gamma_{\nu},
\end{displaymath}
where,
\begin{eqnarray}
\Gamma_{\nu} & = & c_{\nu}^2\left|\langle
0|Q_{\nu}|1\rangle\right|^2\frac{Z_{\nu}(\omega_{\rm
1})}{\omega_{\rm 1}}\coth(\beta\omega_{\rm 1}/2) \nonumber
\\[5pt] & & + \sum_{\gamma>1} (c'_{\nu})^2\left|\langle
0|Q_{\nu}|\gamma\rangle\right|^2
\frac{Z_{\nu}(\omega_{\gamma})}{\omega_{\gamma}}
n(\omega_{\gamma}) \nonumber
\\[5pt] & & + \sum_{\gamma>1}(c''_{\nu})^2\left|\langle 1|Q_{\nu}|\gamma\rangle\right|^2
\frac{Z_{\nu}(\omega_{\gamma}-\omega_{1})}{\omega_{\gamma}-\omega_{1}}
n(\omega_{\gamma}-\omega_{1}). \label{eq:mewidth}
\end{eqnarray}
$Z_{\nu}(\omega)$ is the density of phonon modes of symmetry $\nu$
at energy $\omega$ , $n(\omega) = 1/(e^{\beta\omega}-1)$, and
$\beta=1/k_{\rm B}T$. $c_{\nu}$, $c'_{\nu}$ and $c''_{\nu}$ are
proportional to the magnetoelastic coupling constants and depend
upon the symmetry of the modes.

The corresponding expression for the linewidth due to damping by
spin excitations mediated via an isotropic exchange interaction
$J_{\rm Ex}$ is\cite{Boothroyd-PRL-1996}
\begin{displaymath}
\Gamma_{\rm Ex} = \sum_{\alpha}\Gamma_{\alpha},
\end{displaymath}
where,
\begin{eqnarray}
\Gamma_{\alpha} & = & 2J_{\rm Ex}^2\left|\langle
0|J_{\alpha}|1\rangle\right|^2\chi''(\omega_{\rm
1})\coth(\beta\omega_{\rm 1}/2) \nonumber
\\[5pt] & & + \sum_{\gamma>1}2J_{\rm Ex}^2\left|\langle
0|J_{\alpha}|\gamma\rangle\right|^2 \chi''(\omega_{\gamma})
n(\omega_{\gamma}) \nonumber
\\[5pt] & & + \sum_{\gamma>1}2J_{\rm Ex}^2
\left|\langle 1|J_{\alpha}|\gamma\rangle\right|^2
\chi''(\omega_{\gamma}-\omega_{1})n(\omega_{\gamma}-\omega_{1}).
\label{eq:exwidth}
\end{eqnarray}
Here, $\chi''(\omega)$ is the imaginary part of the local
dynamical susceptibility at the position of the paramagnetic ion.
To model the normal state of the superconductors we take
$\chi''(\omega)\propto\omega$ for reasons discussed in Ref.
\onlinecite{Boothroyd-PRL-1996}.

With Eqs.\ (\ref{eq:mewidth}) and (\ref{eq:exwidth}) written in
this form it is easy to see that there exists a simple mapping
between $\Gamma_{\rm ME}$ and $\Gamma_{\rm Ex}$, as was pointed
out in Ref. \onlinecite{LS-PRB-2000}. Eq.\ (\ref{eq:mewidth})
becomes Eq.\ (\ref{eq:exwidth}) if we replace the $c^2$ constants
by $2J_{\rm Ex}^2$, the $Q_{\nu}$ by the $J_{\alpha}$ operators,
and the function $Z_{\nu}(\omega)/\omega$ by $\chi''(\omega)$.

Having stated the basic formulae I will now proceed to a
comparison of the temperature dependence of the linewidth
predicted by the two models for the case of Ho$^{3+}$ in
YBa$_2$Cu$_3$O$_{6+x}$. For this purpose I will make the same
simplifications as made in Ref.~\onlinecite{LS-PRB-2000}, namely
to use the Debye density of states
$Z(\omega)=3\omega^2/\omega_{\rm D}^2$ with $\omega_{\rm D}=80$
meV for $Z_{\nu}(\omega)$, independent of $\nu$, and to treat all
the magnetoelastic coupling constants as equal and
mode-independent. The quadrupole matrix elements are calculated
from the Ho$^{3+}$ wavefunctions derived from the established
model for the crystal field \cite{Furrer-Staub-Boothroyd}. This
means there is only one unknown parameter $c$ which can be
adjusted to match the experimental linewidth data in the
temperature range over which the superconductor is in the normal
state ($T\gtrsim 100$\,K).

Fig.\ 2(a) displays the curve obtained from the magnetoelastic
model including the first 8 levels of the crystal field split
$J=8$ ground state of Ho$^{3+}$ together with the same data as on
Fig.\ 1. In principle one could include all $2J+1=17$ levels in
the calculation, but we prefer to apply a cut-off after level 8
(11.8\,meV) because there is a large gap to level 9 (at 58\,meV)
at which energy the measured phonon density of states
\cite{Chaplot-PRB-1995} is much smaller than that calculated from
the Debye model. This, together with the small thermal occupancy
at energies of 58\,meV and above, means that the contribution to
the relaxation from levels 9--17 will in reality be negligible. A
comparison of the calculated curves in Fig.\ 1 and Fig.\ 2(a)
reveals a very significant difference between the 3-level and the
8-level approximations. The shape of $\Gamma_{\rm ME}(T)$
calculated in the 3-level approximation has too much curvature in
the temperature range 10--100\,K relative to the more complete
calculation. The reason for this difference, as anticipated
earlier, is the neglect in the 3-level approximation of 5 of the 6
levels in the energy range 1--12\,meV. These levels become
thermally accessible as the temperature is raised and cause a
significant amount of relaxation through indirect transitions.

What is clear from Fig.\ 2(a) is that the experimental points
deviate systematically from the calculated curve for temperatures
below $T\approx100$\,K. This deviation is emphasised in Fig.\ 2(b)
by plotting $\Gamma/\Gamma_{\rm ME}$, the ratio of the
experimental linewidths to the theoretical linewidths derived from
the 8-level magnetoelastic model. Figs.\ 2(a) and (b) can be
compared with the corresponding results for the spin exchange
coupling model shown in Figs.\ 2(a) and (d) of Ref.
\onlinecite{Boothroyd-PRL-1996} which exhibit a similar deviation.

The important point to emphasise, therefore, is that if only 3
levels are included in the relaxation calculation then the
predicted curve follows the experimental points very well, as
illustrated by Fig.\ 1, but as more intermediate levels are
included the experimental linewidths fall significantly below the
calculated curves \emph{both} for the magnetoelastic model
\emph{and} for the spin exchange coupling model (the deviation
being somewhat larger for the latter model). The apparent
agreement  between experiment and theory suggested by Fig.\ 1
cannot, therefore, be used as evidence to favour the
magnetoelastic model over the exchange coupling model as claimed
\cite{footnote-1} by Lovesey and Staub. Indeed, if anything, the
failure of the magnetoelastic model to describe the data depicted
in Figs.\ 2 provides further support for the spin exchange model
because in this scenario the anomalous reduction in linewidth
below $T_c$ is explained by the opening of a gap (the
superconducting and/or pseudogap) in the electronic excitation
spectrum of the CuO$_2$ layers.

The second point I wish to make concerns the magnitude of the
magnetoelastic coupling constant. As indicated on Figs.\ 1 and
2(a) the value of $c^2$ needed to fit the experimental data
changes from 1.93\,meV$^3$ to 0.268\,meV$^3$ as the number of
intermediate levels is increased from 3 to 8. This sensitivity to
the number of levels included in the model is significant if the
coupling constants for different lanthanides are to be compared.
For example, in the magnetoelastic model proposed in Ref.
\onlinecite{LS-PRB-2000} $c$ is proportional to the Stevens factor
$\alpha$ of the lanthanide ion, and so the strength of the
paramagnetic relaxation for one lanthanide ion can be scaled onto
another. In Ref. \onlinecite{LS-PRB-2000} linewidth data for
Tb$^{3+}$ and Ho$^{3+}$ in superconducting YBa$_2$Cu$_3$O$_{6+x}$
were found to be consistent with the predictions of this scaling
hypothesis when analysed with the 3-level model. Mindful of the
sensitivity of $c^2$ to the number of intermediate levels included
in the relaxation calculation I re-analysed the linewidth data
\cite{Staub-JPCM-1999} for Tb$^{3+}$ with a 13-level model and
found $c^2=11.5$\,meV$^3$ (the equivalent value from the 3-level
model is 24\,meV$^3$). My estimate for the ratio $c^2({\rm
Tb})/c^2({\rm Ho})$ is $11.5/0.268=43$, which compares with the
scaling prediction $\alpha^2({\rm Tb})/\alpha^2({\rm Ho})=20.7$, a
factor 2 difference \cite{footnote-2}. Hence, the validity of the
claim in Ref. \onlinecite{LS-PRB-2000} that the scaling of the
linewidths supports the magnetoelastic model for the paramagnetic
relaxation is once again seen to be dependent on how many crystal
field levels are included in the calculation.

I will finish with some brief comments on the two mechanisms,
magnetoelastic or exchange coupling, proposed to explain the
paramagnetic relaxation of lanthanides in YBa$_2$Cu$_3$O$_{6+x}$.
It has not been the intention here to criticize the magnetoelastic
model itself. Indeed, the theory provided by Lovesey and Staub,
summarised in Eq.\ (\ref{eq:mewidth}) above, is an important
contribution that permits progess to be made in the
disentanglement of different sources of relaxation
\cite{footnote-3}. Rather, the purpose of making this Comment is
to illustrate the pitfalls of comparing a minimal model with
experimental data. The inclusion of more relaxation channels has
dramatically changed the quality of agreement with the data, and
one cannot accept the analysis in Ref. \onlinecite{LS-PRB-2000} as
evidence in favour of phonon damping. My opinion is that a true
assessment of whether phononic processes provide an important
mechanism can only be made when more is known about the phonon
density of states and magnetoelastic coupling strengths for
individual phonon modes.

In my opinion one of the strongest arguments for the exchange
coupling interpretation is the evidence in many of the
measurements
\cite{Osborn-PhysicaC-1991,Mukherjee-PRB-1994,Amoretti-PhysicaC-1994,Boothroyd-PRL-1996,Mesot-EPL-1998,Rubio-Temprano-PRL-2000}
for an anomalous reduction in linewidth at a temperature at or
above $T_c$, an effect first observed \cite{Feile-PRL-1981} in the
non-cuprate system La$_{1-x}$Tb$_x$Al$_2$. Such an anomaly cannot
be produced by phonon damping. The absence of any anomaly in other
measurements \cite{Staub-JPCM-1999,Roepke-PRB-1999} can perhaps be
explained by the size of the effect. In most cases where the
crystal field transition probed is below 1\,meV the reduction in
linewidth is found to be typically 0.02\,meV, e.g. Fig.\ 2(a).
This is smaller than the error bar on the data around $T_c$ in
Refs. \onlinecite{Staub-JPCM-1999} and
\onlinecite{Roepke-PRB-1999} which are for quasielastic
transitions, and so it would not have been possible to observe
such an anomaly with the experimental sensitivity available.

It is hoped that the continuing debate over phonon versus spin
fluctuation damping will stimulate a better understanding of
paramagnetic relaxation mechanisms for lanthanide ions in
correlated electron systems, and that the formulae summarised here
may facilitate progress in this area.

%\begin{thebibliography}{10}

%\includegraphics[height=6cm,bbllx=75,bblly=265,bburx=482,
%  bbury=570,angle=0,clip=]{Fig16-paolo-calculation.eps}

\begin{figure}
\begin{center}
\includegraphics[height=6cm,bbllx=40,bblly=200,bburx=534,
  bbury=576,angle=0,clip=]{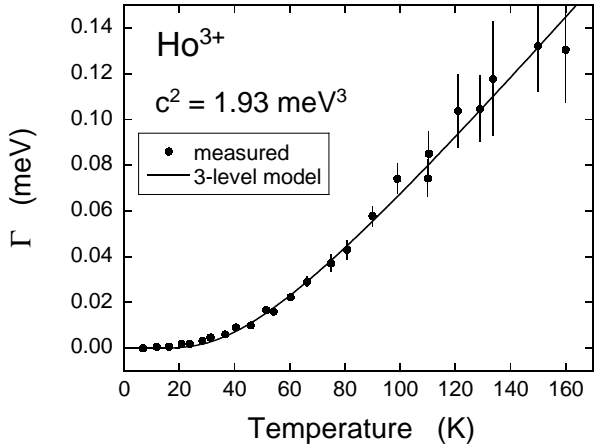}
\vspace{0.3cm}
\caption{Temperature dependence of the intrinsic
linewidth of the ground-state-to-first-excited-state (0.5\,meV)
crystal field transition of Ho$^{3+}$ in
Ho$_{0.1}$Y$_{0.9}$Ba$_2$Cu$_3$O$_7$. The points are the
experimental data from Ref.
\protect\onlinecite{Boothroyd-PRL-1996} and the line is calculated
from the magnetoelastic model of Lovesey and Staub
\protect\cite{LS-PRB-2000} with inclusion of 3 crystal field
levels.} \label{Fig_1}
\end{center}
\end{figure}

\begin{figure}
\begin{center}
\includegraphics[height=12cm,bbllx=173,bblly=339,bburx=402,
  bbury=668,angle=0,clip=]{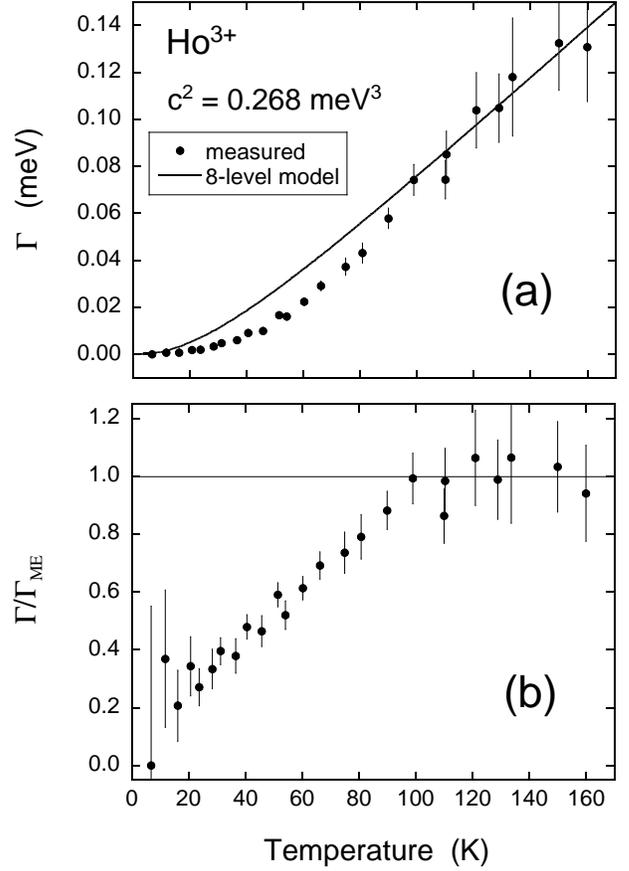}
\vspace{0.3cm} \caption{(a) The same plot as shown in Fig.\ 1 but
with 8 crystal field levels included in the magnetoelastic model
calculation and a reduction in the $c^2$ coefficient from
1.93\,meV$^3$ to 0.268\,meV$^3$. (b) Temperature dependence of the
reduced linewidths obtained by dividing the data by the
theoretical curve in (a).} \label{Fig_2}
\end{center}
\end{figure}


\begin{references}
\bibitem{footnote-2}

Given in addition that the relaxation in Tb$^{3+}$ is dominated by
indirect transitions to a pair of levels near 40\,meV, in contrast
to Ho$^{3+}$ where only levels below 12\,meV are involved, and
that the ratio $Z({\rm 40\,meV})$/$Z({\rm 12\,meV})$ is certain to
be less than that given by the Debye model, the ratio $c^2({\rm
Tb})/c^2({\rm Ho})$ is likely to be still larger.
\end{references}
\end{document}